\documentclass{emulateapj}
\usepackage{apjfonts}

\newcommand{\f}{\phantom{2}}

\newcommand{\ltsimeq}{\raisebox{-0.6ex}{$\,\stackrel 
        {\raisebox{-.2ex}{$\textstyle <$}}{\sim}\,$}} 
\newcommand{\gtsimeq}{\raisebox{-0.6ex}{$\,\stackrel 
        {\raisebox{-.2ex}{$\textstyle >$}}{\sim}\,$}}

\newcommand{\asec}{^{\prime\prime}}
\newcommand{\amin}{^{\prime}}

\def\hst{{\it Hubble Space Telescope~}}

\shorttitle{Imaging of SDSS {\sc z}$>6$ quasar fields}
\shortauthors{Willott et al.}

\begin{document}

\title{Imaging of SDSS {\sc z}$>6$ quasar fields:  gravitational
  lensing, companion galaxies and the host dark matter halos}

\author{
Chris J.\ Willott\altaffilmark{1},
Will J.\ Percival\altaffilmark{2},
Ross J.\ McLure\altaffilmark{2}, 
David Crampton\altaffilmark{1},
John B.\ Hutchings\altaffilmark{1},
Matt J.\ Jarvis\altaffilmark{3},
Marcin Sawicki\altaffilmark{1},
\& Luc Simard\altaffilmark{1}
}

\altaffiltext{\dag}{Based on observations obtained at the Gemini
Observatory, which is operated by the Association of Universities for
Research in Astronomy, Inc., under a cooperative agreement with the
NSF on behalf of the Gemini partnership: the National Science
Foundation (United States), the Particle Physics and Astronomy
Research Council (United Kingdom), the National Research Council
(Canada), CONICYT (Chile), the Australian Research Council
(Australia), CNPq (Brazil) and CONICET (Argentina).}
\altaffiltext{1}{Herzberg Institute of Astrophysics, National Research
Council, 5071 West Saanich Rd, Victoria, B.C. V9E 2E7, Canada;
chris.willott@nrc.ca, david.crampton@nrc.ca, john.hutchings@nrc.ca,
marcin.sawicki@nrc.ca, luc.simard@nrc.ca} 
\altaffiltext{2}{Institute for Astronomy, University of Edinburgh,
Royal Observatory, Blackford Hill, Edinburgh, EH9 3HJ, UK;
wjp@roe.ac.uk, rjm@roe.ac.uk}
\altaffiltext{3}{Astrophysics, Department of Physics, Keble Road,
Oxford, OX1 3RH, UK; mjj@astro.ox.ac.uk}

\begin{abstract}

We have undertaken deep optical imaging observations of three $6.2<z<6.5$
quasar fields in the $i'$ and $z'$ filters. These data are
used to search for foreground galaxies which are gravitationally
lensing the quasars and distant galaxies physically associated with
the quasars. Foreground galaxies are found closer than $5\asec$ from
the lines-of-sight of two of the three quasars. However, the faintness
of these galaxies suggests they have fairly low masses and provide
only weak magnifications ($\mu \ltsimeq 1.1$). No convincing galaxies
physically associated with the quasars are found and the number of
$i'$-band dropouts is consistent with that found in random fields.  We
consider the expected dark matter halo masses which host these quasars
under the assumption that a correlation between black hole mass and
dark matter halo mass exists. We show that the steepness of the
high-mass tail of the halo mass function at this redshift, combined
with realistic amounts of scatter in this correlation, lead to expected
halo masses substantially lower than previously believed. This
analysis can explain the lack of companion galaxies found here and the
low dynamical mass recently published for one of the quasars. 

\end{abstract}


\section{Introduction}

Active supermassive black holes provide a useful means of locating
very distant, massive galaxies. Studies of the host galaxies of
luminous active galactic nuclei (AGN) at redshifts up to $z\approx2$
show they are associated with galaxies with luminous
rest-frame-optical stellar populations, corresponding to $ \gtsimeq
L_{\star}$ (Kukula et al. 2001; Ridgway et al. 2001; Hutchings et
al. 2002; Willott et al. 2003a; Dunlop et al. 2003).  Furthermore, the
strong correlation between black hole mass and stellar bulge
luminosity/mass observed locally (Magorrian et al. 1998; Gebhardt
et al. 2000; Ferrarese \& Merritt 2000) shows that the most massive
galaxies are those whose black holes have accreted the greatest mass
and therefore were the most luminous AGN.

It is now possible to discover luminous quasars out to a redshift of
$z=6.4$. Sources at this redshift are observed as they were $\approx
13$ Gyr ago (93\% of the age of the universe). The Sloan Digital Sky
Survey (SDSS) has made a spectacular breakthrough in locating such
quasars and now 12 are known at $z>5.7$ (Fan et al. 2004). Black hole
masses in these quasars are usually estimated by making the assumption
that the quasars are accreting at the Eddington limit (e.g. Fan et
al. 2001). This method is consistent with a black hole mass
measurement from the kinematics of the broad emission line gas
(Willott, McLure \& Jarvis 2003b). Under the Eddington argument, the
black hole masses of these quasars are mostly in the range $ 1-5
\times 10^9\, {\rm M}_\sun$. These black hole masses could be much
lower in the case of strong gravitational lensing or
beaming. Observations so far have not shown such effects (Fan et
al. 2003; Richards et al. 2004; Willott et al. 2003b).

The space density of these luminous SDSS quasars is extremely low,
$\rho=(6.4 \pm 2.4) \times 10^{-10}\,{\rm Mpc}^{-3}$ (Fan et
al. 2004). The rarity of these objects and their large black hole
masses makes it tempting to associate these quasars with the rarest
peaks in the dark matter density distribution (Fan et al. 2001). The
dark matter halos hosting the quasars would, under this hypothesis,
have a mass of $>10^{13} {\rm M}_{\sun}$.  Such halos would merge with
smaller halos and in the present day would be identified as massive
galaxy clusters with a giant elliptical galaxy hosting the dormant
supermassive black hole at the bottom of the potential well. However,
it is also possible that the SDSS quasars reside in much more common,
lower-mass halos and that the correlation between halo mass and black
hole mass is not well established at this early epoch. Determining the
masses of the quasar host halos is extremely important for
understanding the connection between black hole and galaxy growth in
the early universe.

The evidence from sub-millimetre observations paints a rather
confusing picture. On the one hand, thermal emission from dust has
been detected from several of the quasars. These detections imply huge
dust masses and star formation rates of several thousand ${\rm
M_{\sun}\,yr}^{-1}$ (Bertoldi et al. 2003a; Priddey et al. 2003). At
this rate, the stellar mass of an $L_{\star}$ elliptical could be
built up in $\sim 0.1$ Gyr. The clustering of galaxies with such high
star formation rates at lower redshift ($z\sim3$) is not well
constrained, but early results suggest these galaxies are associated
with massive dark matter halos (Blain et al. 2004, although see
Adelberger 2005). One of these quasars, SDSS\,J1148+5251, has also
been detected in molecular carbon monoxide transitions (Bertoldi et
al. 2003b; Walter et al. 2003). The CO spectra have a relatively
narrow velocity profile with width 280 km${\rm s}^{-1}$. Walter et
al. (2004) obtained high-resolution imaging of the CO emission and
confirm it comes from a compact structure (a few kpc). These
observations allow a dynamical mass estimate for the mass within the
central 2.5\,kpc of the galaxy. This mass is comparable to the
inferred molecular gas mass and is an order of magnitude lower than
the mass predicted by assuming that this quasar resides in one of the
rare peaks corresponding to dark matter halos of
$>10^{13}\,M_{\sun}$. Although there are some uncertainties in the
dynamical mass estimate, particularly the geometry and inclination of
the gas, this result casts serious doubt on the belief that the SDSS
quasars pinpoint the most massive galaxies at high redshift.

Another approach to determining the mass of the quasar host dark
matter halo is via a search for companion galaxies. If the SDSS
quasars are formed in the rarest density peaks, then these correspond
to large-scale overdense regions and hence the number of dark matter
halos located nearby is substantially above the cosmic mean (Kaiser
1984; Barkana \& Loeb 2004). Hence the rarest peaks at $z\approx 6$
are likely to be the sites of the first proto-clusters. If the quasars
really do occupy very massive halos, we would expect to see star
forming galaxies in their vicinity. To attempt to find these companion
galaxies we have carried out deep optical imaging around the three
highest redshift quasars from the sample of Fan et al. (2003). In this
paper we present these data, describe a search for foreground galaxies
which may be gravitationally lensing the quasars and a search for
star forming galaxies at the quasar redshift. Finally, we discuss the
implications for the host halos of the most distant quasars.
Cosmological parameters of $H_0=70\,{\rm km\,s^{-1}\,Mpc^{-1}}$,
$\Omega_{\mathrm M}=0.3$ and $\Omega_\Lambda=0.7$ are assumed
throughout.

\section{Observations}

We present observations for three quasars: SDSS J103027.10+052455.0
(SDSS\,J1030+0524; $z=6.30$), SDSS J104845.05+463718.3
(SDSS\,J1048+4637; $z=6.20$) and SDSS J114816.64+525150.3
(SDSS\,J1148+5251; $z=6.42$).  The three quasar fields were imaged
using the GMOS-North imaging spectrograph on the Gemini-North
Telescope. GMOS-North uses an array of three $2048 \times 4608$ pixel
EEV CCD detectors. The pixel scale is $0.073\asec$ per pixel giving a
useful field-of-view for imaging of $5.5\amin\times5.5\amin$. Since the
typical seeing size of the observations is in the range 0.5 to
$0.7\asec$, the pixels were binned by a factor of 2 in both directions
to speed up readout and data processing. The observations were carried
out in queue mode during November and December 2003. Typical exposure
times are $\approx 2$ hours in the $z'$-band and $\approx 3$ hours in
the $i'$-band. The relative exposure times were designed to give
similar sensitivity in the two bands for very red objects with colours
of $i'-z' \approx 1.5$. More details of the observations are given in
Table\,\ref{tbl-1}.

The GMOS-North detectors have significant fringing at the red end of
the optical wavelength regime. This is of particular concern for the
observations presented here which use the red $i'$ and $z'$ filters.
The fringe pattern remains constant over time, so the fringes can be
successfully removed if one obtains sufficient data to construct a
high signal-to-noise fringe frame free of stars and galaxies. To
enable the construction of these fringe frames, the observations of
each field in each filter were split into 20 to 30 individual frames
and the telescope was dithered in a $3 \times 3$ pattern with offsets
of $10\asec$ in each direction.  Median-combining all the data in each
filter from the three fields enabled the construction of fringe frames
free of astronomical objects.

\begin{deluxetable}{ccccc}
\tablecaption{GMOS-N imaging observations. \label{tbl-1}}
\tablewidth{0pt}
\tablehead{
\colhead{Quasar} & \colhead{Band}   & \colhead{Exposure} & \colhead{$3\sigma$ limiting} & \colhead{Seeing}\\
\colhead{      } & \colhead{    }   & \colhead{time (s)} & \colhead{mag (AB)          } & \colhead{($\asec$)}
}		 
\startdata	 
SDSS\,J1030+0524 & $z'$             & \f 8100             & 26.3                         &   0.68         \\
SDSS\,J1048+4637 & $z'$             & \f 5850             & 26.2                         &   0.61         \\
SDSS\,J1148+5251 & $z'$             & \f 7650             & 26.2                         &   0.65         \\
SDSS\,J1030+0524 & $i'$             & \f 9900             & 27.5                         &   0.65         \\
SDSS\,J1048+4637 & $i'$             &   12150             & 27.7                         &   0.54         \\
SDSS\,J1148+5251 & $i'$             &   13050             & 27.6                         &   0.67         
\enddata
\end{deluxetable}

Reduction of the imaging data was carried out using standard
procedures. Most of the reductions were performed using tasks in the
{\sc IRAF} Gemini package which were specifically designed for
GMOS. The first step is removal of the bias level using a bias
frame constructed from many ($>10$) bias observations (with the same
binning) carried out during the same month. The images were
flat-fielded using flat-field frames generated from observations of
the twilight sky. The three separate CCD images were then mosaiced
together into one large image. The fringe frames generated from all
observations within each filter were then scaled and subtracted from
each image. All images were then inspected and in some cases a
different scaling was applied to better subtract off the fringes.  The
images were then scaled to correct for atmospheric extinction. 

All the images in each filter of the same field were then positionally
registered using detected objects and combined into one image
rejecting pixels deviating by more than 2.5 sigma from the median and
using a bad pixel mask. This method successfully removed CCD defects
and cosmic rays whilst not rejecting counts from actual objects. The
combined images still showed a low-level, large-scale, residual
background which varied across the images. The large-scale background
was successfully removed by applying the background subtraction method
of the Sextractor software (Bertin \& Arnouts 1996) with a mesh size
of $19\asec$. The $i'$ images for each field were then shifted so that
objects would appear in the same locations in the $i'$ and $z'$
images. Astrometry was performed by using the known geometric
distortion of GMOS-North on the sky and setting the positions of the
quasar targets to the positions given in the SDSS survey. Photometric
calibration was achieved using photometric standard stars observed
during the same nights as some of our observations.

\begin{figure}
\resizebox{0.48\textwidth}{!}{\includegraphics{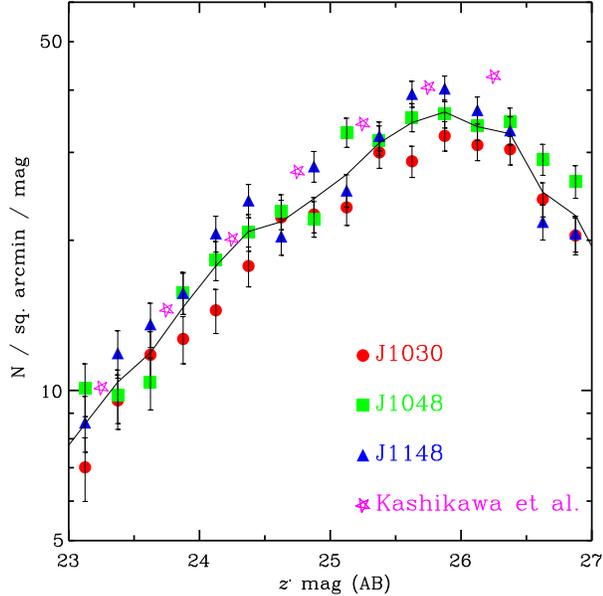}}
\vspace{-0.5cm}
\caption{Binned number counts in the $z'$-band in the fields of the
three $z>6$ quasars. The $z'$-band magnitudes have been
aperture-corrected using the prescription given in
Sec.\,\ref{complete}. Error bars include only Poisson errors. The
solid curve is the mean counts averaged over the three fields. The
open stars show the $z'$-band number counts from the deeper and
wider-area Subaru Deep Field (Kashikawa et al. 2004). The counts are
consistent given the size of the error bars up to $z'=26$. Beyond $z'=26$
the quasar field counts turn over indicating severe incompleteness.
\label{fig:numcts}
}
\end{figure}

\section{Object detection, photometry and completeness}
\label{complete}

Detection of objects in the images was performed using the Sextractor
software. The $z'$-band was selected as the primary detection waveband
since $z>6$ galaxies are expected to have $i'-z'>2$ and may therefore
be undetected at $i'$. Sextractor was run in ``double-image'' mode to
determine $i'$-band measurements for objects detected in $z'$. The
edges of the images do not contain data at all dither positions and
hence have lower sensitivity, contain artifacts and have a varying
background. Objects in these regions were excluded from the object
catalogues. The useful area of the catalogues is 27, 27, 28 square arcmin
for SDSS\,J1030+0524, SDSS\,J1048+4637 and SDSS\,J1148+525,
respectively.

Magnitudes on the AB system were measured in circular apertures of
diameter $1.5\asec$. This size was chosen because it is greater than
twice the seeing size, is much greater than the size scales of known
$z>6$ galaxies (Bouwens et al. 2004) and has lower magnitude errors
than larger apertures. Aperture corrections were applied statistically
to the $z'$-band magnitudes by fitting a linear function to the
difference between the total magnitude and aperture magnitude as a
function of aperture magnitude. The best-fit relation is $z'_{\rm
total} - z'_{\rm ap}=3.58-0.133 z'_{\rm ap}$ which gives an aperture
correction of 0.25 mag at $z'=25$. Note that this relation will lead
to large aperture corrections at bright magnitudes, which may be
inappropriate for compact sources such as stars, but we are in general
only interested in faint objects for which the aperture corrections
are reasonable ($z'>23$). A similar procedure was not adopted for the $i'$-band
magnitudes since the uncorrected aperture magnitudes are used to
measure the $i'-z'$ colours.

The rms noise in the sky background was measured to determine the
magnitude limits of the images. Magnitude limits are quoted as
$3\sigma$ limits in $1.5\asec$ apertures. Typical magnitude limits are
$z'\approx 26.2$ and $i'\approx 27.6$ (see Table\,\ref{tbl-1}). The
relative depths of the images are suitable for detecting very red
objects with $i'-z'>1.5$. 

To assess the completeness of the $z'$-band catalogues we consider
both the observed number counts and the recovery of simulated objects.
Binned number counts for the $z'$-band images for all three fields are
shown in Fig.\,\ref{fig:numcts}. The counts in the three fields do not
differ significantly from each other. They agree well with the
$z'$-band counts determined from a deeper and much larger area survey
(0.2 square degrees or thirty times the GMOS-North field-of-view) of
the Subaru Deep Field by Kashikawa et al. (2004). The number counts in
the quasar fields begin to change slope at $z'>25.5$ and turn over at
$z'=26$, indicating this is where the sample becomes incomplete.

\begin{figure}
\resizebox{0.48\textwidth}{!}{\includegraphics{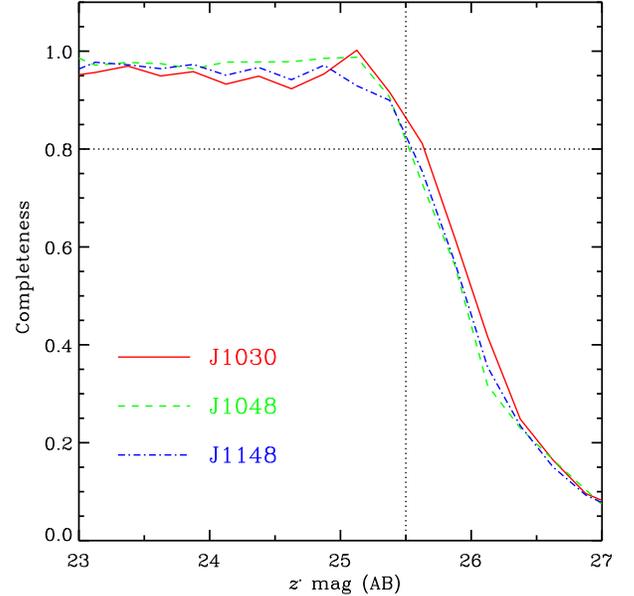}}
\vspace{-0.5cm}
\caption{Completeness ratio against aperture-corrected $z'$-band
magnitude derived from recovery of simulated galaxies as detailed in
Sec.\,\ref{complete}. The curves are quite similar for the three
different fields. Dotted lines indicate the location of a completeness
ratio of 0.8 and the adopted complete magnitude limit of $z'=25.5$.
\label{fig:comp}
}
\end{figure}

The source recovery as a function of magnitude was determined by
populating the images with artificial galaxies and then using
Sextractor to attempt to detect these objects. About 10\,000
artificial galaxies with magnitudes in the range $23<z'<27$ were
placed into copies of the $z'$ images of each quasar. Regions of the
images already occupied by objects were masked out of the process to
eliminate incompleteness due to blending. Sextractor was run twice:
firstly on images containing only the artificial galaxies and a very
low noise level and then on the actual quasar field images with the
artificial galaxies inserted. The ratio of the number of artificial
objects detected in the quasar field images to the number in the low
noise images gives the completeness. This completeness ratio is
plotted as a function of magnitude for the three quasar fields in
Fig.\,\ref{fig:comp}. The completeness in all the fields is fairly
flat at close to 1 up to $z'=25.2$ and then begins to decline. The
rapid decline occurs at $z'>25.5$ and the completeness drops to 0.5 by
$z=26.0$. All the fields have completeness $>0.8$ at $z'=25.5$ and we
adopt this as the magnitude at which completeness begins to become an
issue. This analysis with simulated objects agrees well with the
results for the number counts discussed previously.

\section{Search for foreground galaxies lensing the quasars}

The combined effects of a steep luminosity function and the high
optical depth to $z\sim6$ mean that gravitational lensing is expected
to be particularly important for surveys of luminous quasars at high
redshift, such as the SDSS $z>5.7$ quasar survey. Predictions for the
fraction of multiply imaged quasars in the SDSS for various forms
of the luminosity function have been made by Wyithe \& Loeb (2002a;b)
and Comerford, Haiman \& Schaye (2002). Fan et al. (2003) and Richards
et al. (2004) discuss high-resolution imaging (ranging from $0.1\asec$
to $0.8\asec$) of all seven known $z>5.7$ quasars at the time and
found that none of them appears to be multiply-imaged. However, the
lack of multiple images of the SDSS quasars does not necessarily mean
that lensing is unimportant.  For simple singular isothermal sphere
models the magnification is $\mu<2$ if the source is
singly-imaged. More realistic potentials including isothermal
ellipsoids, cluster-scale halos and micro-lensing can give rise to
higher magnifications for singly-imaged quasars (Keeton, Kuhlen \&
Haiman 2004; Wyithe \& Loeb 2002a).

\begin{figure}
\vspace{0.4cm}
\resizebox{0.48\textwidth}{!}{\includegraphics{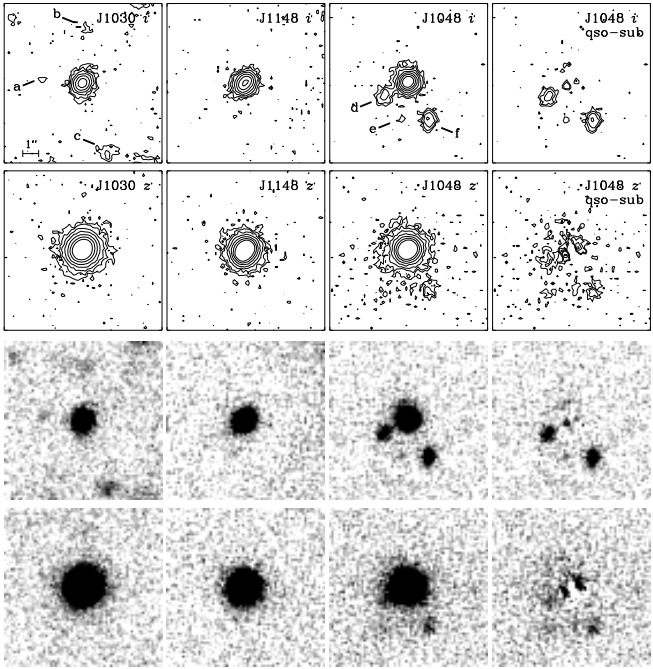}}
\caption{GMOS images in the $i'$- and $z'$-bands centred on the three
quasars SDSS\,J1030+0524, SDSS\,J1148+5251 and SDSS\,J1048+4637 (left
to right). The final column shows the images of SDSS\,J1048+4637 after
subtracting off a scaled point-spread function at the location of the
quasar to show more clearly the nearby galaxies. There are some
residuals from the psf-subtraction process which tests on stars in the
field show are likely to be artifacts.  Each box is $10\asec$
on a side, centred on the quasar with north up and east to the
left. Contour plots are shown with a logarithmic scale at
approximately 1.5,3,5,9,17,31,56\,$\times$ the background rms per
pixel. Greyscale plots have a linear stretch ranging from -0.5\,$\times$
the background rms (white) to +4\,$\times$ the background rms (black).
\label{fig:ims}
}
\end{figure}

Shioya et al. (2002) found a faint galaxy close to the line-of-sight
of the $z=5.74$ quasar SDSS\,J1044-0125. The lensing galaxy has a
magnitude $i'(AB)=24.3$, separation from the quasar $\theta=1.9\asec$
and a likely redshift in the range $1.5<z<2.5$. Spurred on by this
discovery, this group observed two more high-redshift quasar fields
(SDSS\,J1030+0524 at $z=6.30$ and SDSS\,J1306+0356 at $z=5.99$), but
failed to find galaxies projected closer than $6\asec~ {\rm and}~ 3
\asec$ to these quasars, respectively (Yamada et al. 2003).

We have imaged three quasar fields and in this section we use our data
to search for foreground galaxies which may be lensing the
quasars. One of these fields, SDSS\,J1030+0524, has been imaged before
by Yamada et al. (2003), but our images are 1-2 mags deeper and have
better resolution than those in Yamada et al. and Shioya et al. Images
of the regions surrounding the quasars are shown in
Fig.\,\ref{fig:ims}.  Contours and greyscales are shown for the same
fields to highlight different aspects of these high dynamic range
images. The contours show smooth isophotes in the quasar flux which
rule out the existence of very close, bright companion galaxies and
gravitational lensing image splitting in the quasars at the limit of
our resolution. The ellipticities and FWHM of the quasars are
consistent with those of stars in the images.

\begin{deluxetable}{cccccc}
\tablecaption{Objects detected within $5\asec$ of the quasars. \label{tbl-2}}
\tablewidth{0pt}
\tablehead{
\colhead{Object}   & \colhead{Object position (J2000)}   & \colhead{$\theta / \asec$}    & \colhead{$i'$ (AB)} & \colhead{$z'$ (AB)}
}		 
\startdata	 
a & 10:30:27.26 +05:24:55.3  &   2.5 &   $27.54\pm0.35$ &  $ >26.3$         \\
b & 10:30:27.08 +05:24:58.2  &   3.5 &   $26.69\pm0.17$ &  $26.12\pm0.28$   \\
c & 10:30:26.99 +05:24:50.6  &   4.6 &   $26.05\pm0.10$ &  $ >26.3$         \\
d & 10:48:45.19 +46:37:17.4  &   1.7 &   $25.67\pm0.05$ &  $24.97\pm0.11$   \\
e & 10:48:45.10 +46:37:16.1  &   2.2 &   $27.32\pm0.22$ &  $ >26.2$         \\
f & 10:48:44.92 +46:37:15.9  &   2.7 &   $25.37\pm0.04$ &  $24.82\pm0.10$   
\enddata
\tablecomments{Objects detected at $i'$-band within a $5\asec$ radius
of the quasars.  Three objects were detected in each of the fields of
SDSS\,J1030+0524 and SDSS\,J1048+4637. Nothing was detected within
$5\asec$ of SDSS\,J1148+5251. Object refers to the labels in
Fig.\,\ref{fig:ims}. $\theta$ is the separation of the object and the
quasar.  Magnitudes are given as observed in a $1.5\asec$ diameter
aperture. Limits at $z'$-band are $3\,\sigma$ in the same size
aperture.}
\end{deluxetable}

We searched for objects close to the quasars by inspection of smoothed
images. We only consider objects within a $5\asec$ radius of the
quasars, since lensing by galaxy-sized masses is only effective with
impact parameters of a few arcsec. In addition, the probability of
finding a galaxy at random within a circle of $5\asec$ radius is large
at the magnitude limit of the $i'$-band images (this probability is
0.9 for $i'<27$).

Three galaxies were detected in the fields of each of SDSS\,J1030+0524
(labelled a, b, c) and SDSS\,J1048+4637 (d, e, f) and none in the
field of SDSS\,J1148+5251. The details of these galaxies are given in
Table\,\ref{tbl-2}. All of these galaxies are visible in the $i'$-band
and their $i'-z'$ colours (or colour limits) show they are foreground to the
quasars. For SDSS\,J1030+0524, the objects are all rather faint
($i'>26$) and distant from the quasar ($\ge 2.5\asec$). For
SDSS\,J1048+4637, the closest galaxy (d) is only $1.7\asec$ from the
quasar. To measure accurate photometry for this galaxy it was
necessary to subtract the bright unresolved quasar from the image
(far-right panels of Fig.\,\ref{fig:ims}). Galaxy d has $1.5\asec$
aperture magnitudes of $i'=25.67\pm0.05$, $z'=24.97\pm0.11$. The
galaxy $2.7\asec$ from the quasar (f) has similar magnitudes. Both
these galaxies have fairly red colours with $i'-z'=0.70$ and $0.55$,
respectively. Only 20\% of galaxies in our images with similar
magnitudes have $i'-z'>0.5$.

\subsection{Lensing galaxy masses, redshifts and magnifications}

In order to determine the lensing magnification due to such galaxies
one needs to know the galaxy mass (or equivalently the velocity
dispersion) and redshift. Under the simplifying assumption that the
lens potential can be modelled as an isothermal sphere the
magnification $\mu$ is $\mu = \theta / (\theta - \theta_{\rm E})$,
where $\theta$ is the angle from the lens galaxy to the source,
$\theta_{\rm E}$ is the Einstein radius given by
$\theta_{\rm E} = 4 \pi (\sigma_v/c)^2(D_{\rm LS}/D_{\rm OS})$,
where $D_{\rm LS}$ is the lens--source angular diameter distance and
$D_{\rm OS}$ is the observer--source angular diameter distance.

Neither the masses nor redshifts are strongly constrained by our
photometric observations, but we will use the information available to
determine likely values. The red colour of $i'-z'=0.70$ for galaxy d
close to SDSS\,J1048+4637 is not consistent with the colour of known
galaxies at low redshifts. Therefore we can at least put a lower limit
on the redshift of this galaxy. Also, galaxy colours at high redshifts
are expected to be blue in the absence of dust reddening, due to the
prominence of young stellar populations. We used the HyperZ code
(Bolzonella, Miralles \& Pell\'o 2000) with the 2 photometric points
to determine if this colour leads to any constraints on the
redshift. Using a range of evolving galaxy templates with moderate
dust reddening ($0 \leq A_V \leq0.5$) we find a 90\% confidence range
for the redshift is $0.4<z< 2.1$. There is also a secondary solution
at $5.1<z<5.5$ where the red colour is due to the Lyman break entering
the $i'$-band. Considering the redshift distribution of galaxies
of this magnitude it is most plausible that the galaxy is at
the lower redshift range. Note that if this galaxy is {\em highly}
reddened by dust, then the redshift is no longer well-constrained.

The mass or stellar velocity dispersion of the galaxy can be estimated
from the photometry by assuming a redshift and star formation history.
Given the red colour of this galaxy we assume that it is an early-type
which has undergone passive evolution after a starburst at high
redshift and use the evolving elliptical galaxy template from Bruzual
\& Charlot (2003) which formed at $z=7$ (the exact formation redshift
assumed does not critically change the results presented here). The
correlation between $i'$-band absolute magnitude and $\sigma_v$
observed at low redshifts (Bernardi et al. 2003) was used to relate
the evolved absolute magnitude to $\sigma_v$. For three possible lens
redshifts of $z=0.4,1,2$ the calculated stellar velocity dispersion is
$\sigma_v=55, 87, 127\,{\rm km\,s}^{-1}$. Note that the Bernardi et
al. relation is only derived for $z<0.3$ and $\sigma_v>100\,{\rm
km\,s}^{-1}$, so its use here involves extrapolation in both redshift
and velocity dispersion.  Given the redshift of the source is $z=6.20$
and $\theta=1.7\asec$ this leads to Einstein radii $\theta_{\rm
E}=0.07, 0.13, 0.18\asec$ and magnifications $\mu=1.04, 1.08,1.12$ for
lens redshifts of $z=0.4,1,2$, respectively.

It is clear that, whatever the redshift of this lensing galaxy, it is
not very massive and provides only a small magnification to the flux
of SDSS\,J1048+4637. Therefore it makes a negligible difference to the
derived luminosity and black hole mass. Given that the closest galaxy
to SDSS\,J1030+0524 (galaxy a) is even fainter and further away, the
magnification in that case will be even lower. An increase in the
magnification could come about if (i) the galaxy lies in a moderately
rich high-redshift cluster; or (ii) the galaxy resides in a dark
matter halo with higher than average mass-to-light ratio.

For SDSS\,J1044-0125, Shioya et al. (2002) estimated the lens galaxy
to have a velocity dispersion of $\sigma_{v}\sim200$\,km\,s$^{-1}$ and
be located at redshift $z\approx 2$.  According to Shioya et al. this
would give a magnification of $\mu=2$. However we have repeated their
calculations and find their values for the Einstein radius and
magnification to be erroneous (this error is also present in Yamada et
al. 2003). For $z=2$ and $\sigma_{v} =200$\,km\,s$^{-1}$, the
magnification is actually $\mu=1.3$. The range of plausible lens
redshift and velocity dispersion combinations given in their paper
lead to a range of $1.1<\mu<1.5$. As with SDSS\,J1048+4637, the
corrections to the derived luminosity and black hole mass of
SDSS\,J1044-0125 are small.

\begin{figure}
\resizebox{0.48\textwidth}{!}{\includegraphics{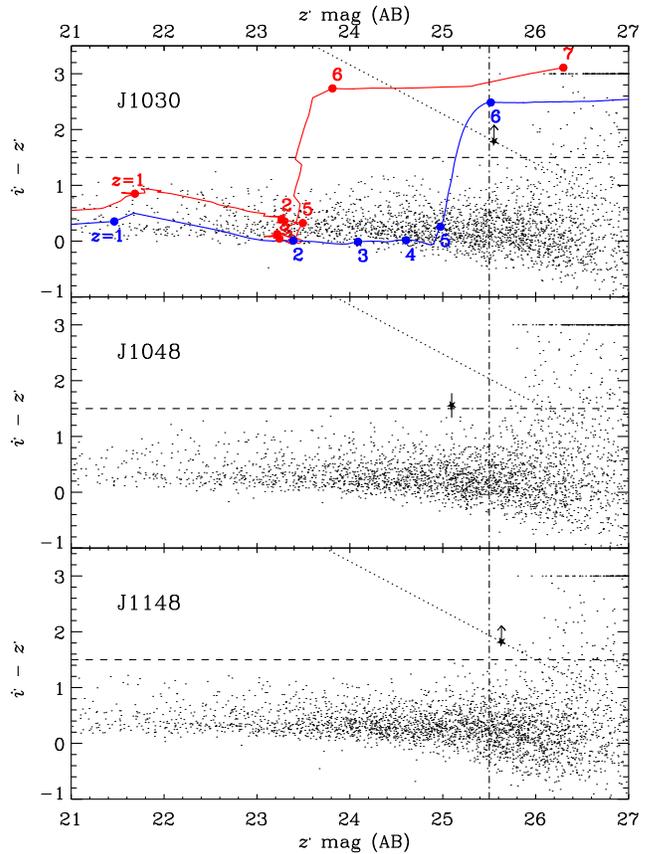}}
\caption{Colour-magnitude (aperture-corrected) diagrams for objects
detected in $z'$ in the quasar fields. The completeness limit of
$z'=25.5$ is shown with a dot-dashed line. The high-redshift galaxy
selection criterion of $i'-z'>1.5$ is marked with the dashed line. The
dotted line represents the colour of an object which is just detected
at the $3\sigma$ level in the $i'$-band as a function of the $z'$-band
magnitude.  Most objects with measured zero or negative flux in the
$i'$-band are plotted at $i'-z'=3$. The exceptions to this are objects
which pass the $i'$-band dropout selection criteria discussed in
Sec.\,\ref{search}.  These are shown with star symbols and lower
limits on the $i'$-band $3\sigma$ line. The dots at $i'-z'>1.5$
indicate sources which are detected at less than $4\sigma$ at
$z'$-band and hence have very large uncertainties on their magnitudes
and colours and in some cases may be spurious.  The labelled curves
show the colour and magnitude as a function of redshift for an
evolving L$^{\star}$ elliptical (upper curve) and a non-evolving
L$^{\star}$ Lyman-break galaxy (lower curve) -- see text for more
details.
\label{fig:colmags}
}
\end{figure}

\section{Search for $z>6$ galaxies in the quasar fields}
\label{search}

Galaxies at redshifts $z>5.7$ can be identified by the sharp drop in
flux across the Lyman break leading to a very red $i'-z'$ colour. The
$z>6$ SDSS quasars have $z'\approx 20$ and colours measured from our
images of $i'-z'=3.25,2.98,3.25$ respectively for SDSS\,J1030+0524,
SDSS\,J1048+4637 and SDSS\,J1148+5251.  The spectra of quasars and
star forming galaxies over the rest-frame wavelength range
$90-140$\,nm probed by the $i'$ and $z'$ filters are dominated by a
large break due to absorption by neutral hydrogen. Therefore one would
expect companion galaxies to have comparable $i'-z'$ colours to the
quasars. The fact that the light from a companion galaxy would pass
through IGM neighbouring that which the quasar light passes through,
further strengthens the idea that the colours of companion galaxies
are expected to be $i'-z'\approx 3$.

Simulated colour-magnitude tracks for two different types of galaxy as
a function of redshift are shown on Fig.\,\ref{fig:colmags}. Model
galaxy spectra were generated from the Bruzual \& Charlot (2003)
spectral synthesis code with Lyman forest absorption evolution
matching the observations of Songaila \& Cowie (2002). The upper curve
is a present-day L$^{\star}$ elliptical which formed all of its stars
in a starburst starting at $z=10$ with a characteristic timescale of
1\,Gyr and evolved since without merging. There is no dust extinction
assumed for this model, but in reality the dust extinction would
increase with redshift (due to evolution and $k$-correction) making
the galaxy redder and fainter than plotted at higher redshifts. The
lower curve is a L$^{\star}$ Lyman-break galaxy model where the galaxy
is observed 0.5\,Gyr into a constant star formation rate starburst.
It is clear that the only possible low-redshift contaminants with
$i'-z'>1.5$ are galaxies with a high level of dust reddening.

Also plotted on Fig.\,\ref{fig:colmags} are colour-magnitude diagrams
constructed from the $z'$-band-selected catalogues described in
Sec.\,\ref{complete} for each of the three quasar fields. Most objects
have colours in the range $0<i'-z'<1$ as is well known from previous
surveys (Dickinson et al. 2004; Capak et al. 2004). These colours are
in agreement with the model galaxy tracks at $z<5$.

A search was made for objects which could plausibly be high-redshift
galaxies. The $i'$-band dropout selection criteria adopted were
signal-to-noise ratio in the $z'$-band $\ge 4$ and a colour of
$i'-z' \ge 1.5$. Possible candidates were inspected and magnitudes
checked to ensure their unusual colours are not spurious. A total of
three objects satisfying these criteria were found - these are shown
with filled symbols on Fig.\,\ref{fig:colmags}. These all have
magnitudes in the range $25<z'<26$ and only one of them is detected at
the $3\sigma$ level at $i'$-band. This object has $z'=25.10$ and a
colour of $i'-z'=1.56 \pm 0.22$. The other two objects have $z'
\approx 25.6$ and only lower limits on their $i'-z'$ colours which lie
in the range $1.7-1.9$.

In Sec.\,\ref{complete} we showed that our $z'$-band images are
complete at the greater than 80\% level to $z'=25.5$.
Fig.\,\ref{fig:colmags} shows that at $z' \le 25.5$, every single
$z'$-band object detected on our images has a counterpart at the
$>3\sigma$ level in the $i'$-band image. At fainter magnitudes, this
is no longer true and our constraints on the $i'-z'$ colours of
objects at these magnitudes becomes very weak due to uncertainty in
both the $z'$ and $i'$ magnitudes. Therefore we will limit further
analysis to objects brighter than the magnitude limit of $z'=25.5$.

At $z'<25.5$ we find one $i'$-band dropout in the three quasar
fields. This object has $i'-z'=1.56 \pm 0.22$ which is quite close to
the colour selection value and the size of the uncertainty means it is
quite plausible that photometric errors have scattered the colour into
the dropout range. The measured colour is about $7\sigma$ away from
the colours of the SDSS quasars, which suggests that if it is a
high-redshift galaxy, then it is most likely foreground to the quasars
with a redshift in the range $5.7<z<6$. The galaxy is located a
projected distance of $81\asec$ from the quasar (approximately
halfway from the quasar to the edge of the GMOS field). Therefore we
conclude that there are no plausible galaxies brighter than $z'=25.5$
associated with any of these three quasars.

We now consider the number of $i'$-band dropouts we could have
expected to find under the assumption that the quasar fields are
``random'' and show no enhancement due to the existence of the
quasars. The best comparison datasets which go deep enough over a wide
area are the {\it Hubble Space Telescope} ACS imaging of the GOODS
regions (Giavalisco et al. 2004) and the Subaru Deep Field (Kashikawa
et al. 2004). These observations give a surface density of objects
with $z'<25.5$ and $i'-z'>1.5$ of $0.01-0.02$ arcmin$^{-2}$ (Dickinson
et al 2004; Bouwens et al. 2004; Nagao et al. 2004). The total sky
area we have surveyed with GMOS-North is 82 arcmin$^{2}$.  Therefore
on the basis of the GOODS and SDF observations we would expect
$\approx 1-2$ $i'$-band dropouts in our total area. Our finding of one
dropout is entirely consistent with the expectations for a blank
field.

The results presented in this section show that these quasar fields do
not exhibit an excess of luminous companion galaxies. The magnitude
limit of $z'=25.5$ corresponds to a UV luminosity of
$L_{1500}=2.5\times10^{29}{\rm erg s}^{-1}{\rm Hz}^{-1}$ at a redshift
of $z=6.3$. This is equivalent to 2\,L* in the $z\approx6$ galaxy luminosity
function (Bunker et al. 2004) and an unobscured star formation rate of
$SFR=30\, {\rm M}_\sun {\rm yr}^{-1}$, assuming the conversion given
in Madau, Pozzetti \& Dickinson (1998). The few known galaxies at
redshifts $z\approx 6.6$ discovered in narrow-band surveys have star
formation rates derived from their UV luminosities comparable to this
limit (Hu et al. 2002; Kodaira et al. 2003). For comparison, the
millimeter detections of dust in SDSS\,J1048+4637 and SDSS\,J1148+5251
imply total star formation rates $> 1000\, {\rm M}_\sun {\rm yr}^{-1}$
in the host galaxies of the quasars (Bertoldi et al. 2003a).

A similar study to ours has recently been reported by Stiavelli et
al. (2005). They have observed the quasar SDSS\,J1030+0524 with ACS on
the \hst in the $i_{775}$ and $z_{850}$ filters. Their data reaches
slightly deeper than ours and they identify seven objects with
$i_{775}-z_{850}>1.5$ in the quasar field. Only 3\% of
randomly-selected areas in the GOODS fields show such a high density
of red objects, indicating an excess in the field of this quasar.  The
single $i'-z'>1.5$ object we discovered in the field of
SDSS\,J1030+0524 which is plotted on Fig.\,\ref{fig:colmags} is also
found by Stiavelli et al. and a high-resolution image of it is shown
in the upper panel of their Fig.\,2. Four of the seven objects found
with ACS are detected at $i_{775}$ and have $i_{775}-z_{850}<2$. These
objects are not red enough to be located at a redshift similar to the
quasar (Dickinson et al. 2004). A spectroscopic redshift has been
obtained for one object and it reveals a redshift of $z=5.970$. The
large redshift difference between the quasar and this galaxy means
that if they are part of the same large-scale overdensity then the
comoving scale of that structure is larger than anything found in the
low-redshift universe (Pandey \& Bharadwaj 2005).

\section{Constraints on the dark matter halo masses}

Early-type galaxies in rich clusters in the local universe show
homogeneous, old stellar populations suggesting coeval star formation
at high redshifts (e.g. L\'opez-Cruz, Barkhouse \& Yee 2004).
Therefore, if the quasars (two of which have far-IR luminosities
indicating vigorous star formation) are in proto-cluster environments
one would expect to see massive companion galaxies forming stars.  It
has been suggested that there could be a suppression of star formation
in the vicinity of a strong UV photo-ionizing field such as that of a
quasar (Couchman \& Rees 1986). This is because the radiation would
heat the diffuse gas in the IGM to a temperature of $T\sim 1000$\,K
and inhibit the cooling process by which gas condenses to form
stars. However, these effects are likely to be significant only in low
mass halos and will not be such an issue for the more massive halos
hosting galaxies with $SFR>30\, {\rm M}_\sun {\rm yr}^{-1}$ which we
are sensitive to.

The goal of our observations was to constrain the masses of the dark
matter halos hosting these quasars by considering the clustering of star
forming galaxies around them. In fact, we have been unable to identify any
companion galaxies suggesting that these quasars may not be residing in
the most massive halos at this epoch, in line with the dynamical mass
measurement of the host galaxy of SDSS J1148+5251 by Walter et al. (2004).
However, given the large uncertainties in transforming from observable
quantities such as UV star formation rate to typical Lyman break halo mass
and duty cycle at $z>6$, it is difficult to directly turn our
non-detection into a robust upper limit for the halo masses. In this
Section we show that scatter in the $M_{BH}-M_{DM}$ relation, combined
with a steeply falling mass function, biases the host halo masses of the
observed high-redshift quasars to lower values, suggesting that the
inferred low host halo masses do not necessarily imply a break-down of the
low redshift $M_{BH}-M_{DM}$ correlation at $z>6$.

An upper limit on the halo mass hosting luminous quasars can be
calculated from a comparison of their space densities. This method
assumes that quasars reside in the most massive halos which exist in
sufficient quantity to host all the observed quasars. The nine
luminous quasars at redshift $5.7<z<6.5$ discovered by the SDSS (which
all have black holes with mass $M_{\rm BH}=1-4 \times 10^9\, {\rm
M}_\sun$ via the Eddington argument) give a space density of $(6.4 \pm
2.4) \times 10^{-10}\, {\rm Mpc}^{-3}$ (Fan et
al. 2004)\footnotemark. The halo mass function at $z=6.3$ was
determined from the mass function fit to the numerical simulations of
Sheth \& Tormen (1999). The quasar space density corresponds to the
space density of all halos with $M_{\rm DM}>1.5\times10^{13}\,M_\sun$. This
limit is cosmology dependent, and is obviously strongly dependent on
the high mass tail of the Sheth \& Tormen mass function, which is only
relatively sparsely sampled in the simulations used to determine this
fit. However, in support of this functional form, Barkana \& Loeb
(2004) show that the tail provides a good approximation to the
numerical mass function at extremely high redshifts, provided that the
lack of large scale modes in the simulations is taken into account.

\footnotetext{We neglect the likely existence of quasars just as
powerful as those in the SDSS which have their UV luminosities
substantially decreased by dust. The space density of obscured quasars
at lower redshifts is comparable to that of unobscured quasars (Zheng
et al. 2004). However the ratio of obscured to unobscured quasars at
$z=6$ is so uncertain that we do not attempt to make any correction to
the quasar space density.}

An estimate of the host halo masses can be found by using the
estimated black hole masses and further assuming that there is a
correlation between black hole and halo masses. Such a correlation has
been shown to exist at low redshifts (Ferrarese 2002) via a
combination of the well known stellar bulge--black hole mass
correlation and a correlation between halo circular velocity and bulge
velocity dispersion. The correlation determined by Ferrarese (2002)
with a correction applied by Bromley et al. (2004) is:
\begin{equation}
\frac{M_{\rm BH}}{10^8\, {\rm M}_{\sun}} \sim 0.015\left(\frac{M_{\rm
DM}}{10^{12}\, {\rm M}_{\sun}}\right)^{1.82}.
\end{equation}
The universality of this correlation is untested and in particular
whether it evolves with redshift is highly uncertain. We will refer to
this non-evolving correlation between $M_{\rm BH}$ and $M_{\rm DM}$ as
case A.

Wyithe \& Loeb (2003) have noted that in feedback-regulated black hole
growth models (e.g. Silk \& Rees 1998; Fabian 1999) the important
parameter for the strength of the dark matter halo potential is the
circular velocity, not the mass. The relationship between circular
velocity and mass is redshift-dependent due to the universal
expansion: $v_c \propto M_{\rm DM}^{1/3}\,(1+z)^{1/2}$ (e.g Barkana
\& Loeb 2001). The feedback-regulated model of Wyithe \&
Loeb\footnotemark\, leads to a redshift-dependent correlation between
$M_{\rm BH}$ and $M_{\rm DM}$ of
\begin{equation}
\frac{M_{\rm BH}}{10^8\, {\rm M}_{\sun}} \sim f\left(\frac{M_{\rm
DM}}{10^{12}\, {\rm M}_{\sun}}\right)^{5/3}(1+z)^{5/2}.
\end{equation}
Since the dormant black holes observed in the local universe typically
formed at redshifts $1<z<2$ (e.g. Yu \& Tremaine 2002) we fix the
normalizing factor $f$ so that the correlation roughly matches that of
Bromley et al. (2004) at $z=1.5$. This condition is met by $f=0.0018$.
We will refer to this evolving correlation between $M_{\rm BH}$ and
$M_{\rm DM}$ as case B. 

\footnotetext{We omit the parameter $[\xi(z)]^{5/6}$ from the equation
  given in Wyithe \& Loeb (2003) since it is close to unity and
  negligible compared with the uncertainties involved.}

Since the two possible scenarios are normalized to each other at
$z=1.5$, at higher redshifts the host halo for a black hole of a given
mass will be less massive for case B than case A. Specifically, for a
quasar at $z=6.3$ with $M_{\rm BH}=2 \times 10^9\, {\rm M}_\sun$, then
case A gives $M_{\rm DM}=5.2 \times 10^{13}\, {\rm M}_\sun$ and case B
gives $M_{\rm DM}=1.3 \times 10^{13}\, {\rm M}_\sun$.  A simple
comparison of the case A estimate with the space-density limit shows
that there are insufficient halos with masses of $M_{\rm DM}=5 \times
10^{13}\, {\rm M}_\sun$ at these high redshifts to host the number of
observed quasars. The simple conclusion to draw is that the
correlation between black hole mass and halo mass does not remain the
same at $z>6$ as it is locally. A similar conclusion was reached by
Bromley et al. (2004) by considering the total black hole mass density
built up by quasars. Wyithe \& Loeb (2005) showed that the evolution
in the 2dF quasar correlation function over the (admittedly narrow) range
$1<z<2$ was more consistent with the evolving $M_{\rm BH}-M_{\rm DM}$
correlation than the redshift-independent one. In what follows we will
use both case A and B correlations, whilst remembering that prior
evidence supports case B over A.

An important ingredient in the application of such correlations that
is often overlooked is the effect of intrinsic scatter in the
relation. We now incorporate the scatter in these correlations to
estimate the expected host halo masses. We will show that including
the scatter makes a significant difference due to the steepness of the
halo mass function for these rare halos. This is because lower mass
halos are much more abundant, so low mass halos which are outliers
in the $M_{\rm BH}-M_{\rm DM}$ correlation could be more abundant than
high mass halos on the correlation.

We begin with the dark matter halo mass function fit of Sheth \&
Tormen (1999). This is evaluated at a redshift of $z=6.3$. Above a
mass of $\sim 10^{12}\, {\rm M}_\sun$, the mass function is
exponentially declining. The black hole mass within each halo is given
by the case A or B correlations. We now add scatter to these black
hole masses. The intrinsic scatter in the correlation between black
hole mass and stellar velocity dispersion at low redshift is in the
range $0.25-0.30$ dex (Tremaine et al. 2002). The conversion from
stellar velocity dispersion to halo mass is likely to add in
considerably more scatter, so we consider 0.4 dex to be the minimum
amount of scatter in this correlation at low redshift. This scatter
may increase substantially at higher redshifts, particularly for
galaxies which are in the process of growing their black holes or
accreting matter from the surrounding IGM. To bracket the uncertainty
in the scatter we consider four different cases with 0.4, 0.6, 0.8 and
1.0 dex scatter in the correlation. Although the shape of the scatter
is unknown even locally, we will adopt a lognormal distribution as the
simplest shape.

\begin{figure}
\resizebox{0.48\textwidth}{!}{\includegraphics{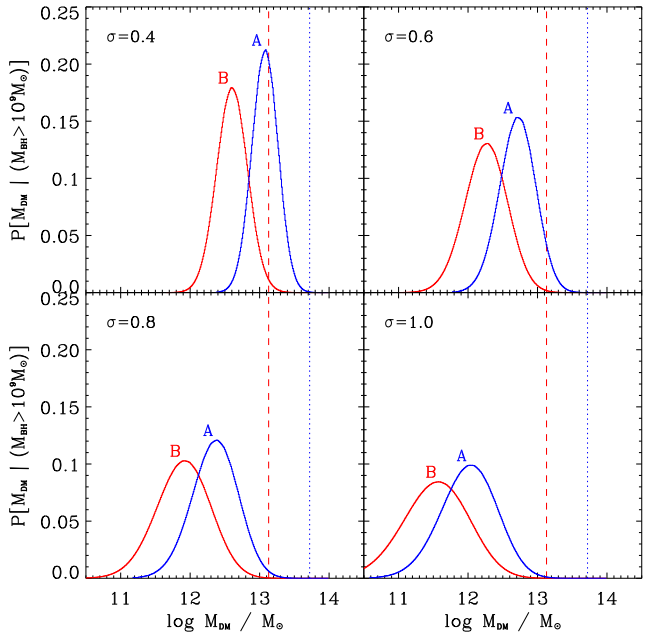}}
\caption{Probability distribution for the mass of the dark matter halo
hosting a black hole with mass $>10^9\, {\rm M}_{\sun}$ at $z=6.3$.
These curves were generated assuming correlations between $M_{\rm BH}$
and $M_{\rm DM}$ of the form A and B (see text) with lognormal
intrinsic dispersion $\sigma$. The vertical dotted (dashed) line shows
the case A (B) $M_{\rm DM}$ expected for $M_{\rm BH}=2\times10^9\,
{\rm M}_{\sun}$ without considering the combined effects of scatter
and the steepness of the halo mass function. These plots clearly show
that the expected halo masses are considerably lower than obtained by
simply using the $M_{\rm BH}-M_{\rm DM}$ correlation without scatter.
\label{fig:mbhmdm}
}
\end{figure}

By identifying the most luminous $z\sim6$ quasars across a large
fraction of the sky, the SDSS has selected the most massive, active
black holes at this epoch (under the assumption that luminous quasars
at such early times are accreting at the Eddington limit). Therefore
we take a black hole mass cut of $M_{\rm BH}> 10^{9}\, {\rm M}_\sun$
and then calculate the probability distribution of dark matter halos
which host such massive black holes given the correlations A and B and
the various amounts of scatter in the correlation. The results are
shown in Fig.\,\ref{fig:mbhmdm}. Including scatter has considerably
decreased the expected masses of the halos hosting these quasars. Even
for the case with minimum scatter there is a shift of $\sim 0.5$ in
$\log_{10}M_{\rm DM}$. For the maximum scatter case considered, this
shift in $\log_{10}M_{\rm DM}$ is $\sim 1.5$. 

Walter et al. (2004) use resolved CO emission to measure a dynamical
mass for the inner 2.5 kpc of SDSS\,J1148+5251 to be $\sim
5\times10^{10}\, {\rm M}_\sun$. They argued this to be at least a
factor of 10 lower than expected by naive extrapolation of the low
redshift correlation between black hole mass and stellar velocity
dispersion. We have shown here that using the case B
redshift-dependent correlation and/or including the effects of a
reasonable amount of scatter ($\sigma \gtsimeq 0.6$) lead to an
expected halo mass at least an order of magnitude lower than using the
naive correlation of case A. This provides a simple explanation for
the observations of Walter et al.  and those presented in this paper,
without the need for the $M_{\rm BH}-M_{\rm DM}$ correlation to break
down at high redshift. A consequence of this is that the SDSS quasars
may not be identifying the most overdense regions of the high-$z$
universe as is commonly assumed.

\section{Summary}

We have presented deep imaging in the $i'$ and $z'$ filters of the
fields of three of the most distant known quasars. For two of the
three quasars, there are foreground galaxies at projected distances of
a few arcsec. However, the faintness of these galaxies suggests they
have relatively low masses which, combined with the large impact
parameters, provide only a weak gravitational lensing magnification
that would not significantly alter the derived quasar luminosities.

A search for star forming galaxies at the redshifts of the quasars was
carried out. No enhancement in Lyman-break galaxies above that in
random fields is found. The UV continuum star formation limit reached
is $30\, {\rm M}_\sun {\rm yr}^{-1}$, comparable with known $z\approx
6.5$ galaxies. The lack of companion galaxies could be indicative of
these quasars residing in lower mass dark matter halos than previously
believed. We consider the effect of scatter in correlations between
black hole mass and halo mass and find that this scatter means that we
expect the most massive black holes to be found in a range of halo
masses, typically much lower than found by use of a simple correlation
between the two quantities.

To obtain a more accurate estimate of any overdensity of star forming
companions will require much deeper data. Whilst few-orbit ACS
observations are able to find a few possible companions (Stiavelli et
al. 2005), a proper census requires something of similar depth to the
Hubble Ultra Deep Field (reaching 0.1 L* in the $z=6$ galaxy
luminosity function with a limiting magnitude of $z'=29$; Bunker et
al. 2004) and such observations will therefore require a large
increase in sensitivity such as from the {\it James Webb Space
Telescope}.

\acknowledgments

Thanks to the Gemini-North queue observers for executing the
observations presented here. We thank Stuart Wyithe and Meghan Gray
for interesting discussions. Thanks to the anonymous referee for some
useful suggestions to improve the paper.

\end{document}